\documentclass[aps,groupedaddress,prd,preprint,a4paper]{revtex4}
\usepackage{graphicx}
\usepackage[dvips]{pstcol}
\usepackage{pst-node}
\usepackage{pst-plot}
\usepackage{pst-coil}
\begin{document}
\begin{center}
%\draft
\title{Single Top Production at HERA and THERA}
\author{A.T.ALAN and A.SENOL}
\address{Department of Physics, Faculty of Sciences and Arts, Abant Izzet Baysal University,14280 G\"{o}lk\"{o}y, Bolu, TURKEY}
\date{\today}
\begin{abstract}

We study the single top production mediated by flavor changing
neutral current via both of the $t-q-\gamma$ and $t-q-Z$ vertices
(here q represents c and u quarks) in $ep$ collisions at two
colliders HERA and THERA. Contribution of the second vertex
becomes even more important as the couplings take more improved
values provided by the higher luminosities of colliders. In
addition to these improvements if the CM energy of the collider is
increased, the production will be  dominated by the anomalous
$t-q-Z$ vertex.

\end{abstract}
\maketitle
\end{center}
\section{Introduction}
It is well known that within the Standard Model (SM), single top
quark production is negligible because of the flavor conservation
in the leading order.However in a wide class of dynamical models
extending the SM this possibility exists due to the large top mass
close to the electroweak symmetry breaking scale. Flavor Changing
Neutral Current (FCNC) interactions of top quarks has a particular
importance. In these models the top quarks is predicted to have
significantly large FCNC interactions \cite{HF,THPZ}. Production
of single top quarks via FCNC vertices was extensively studied at
$e^+e^-$ \cite{THJL,OSY} and at hadron colliders \cite{TC,TC2,MT}
. At HERA, as a lepton-hadron type collider, the same production
was first investigated in \cite{HFD}. In this reference only the
charm distribution in the proton is considered. The H1
Collaboration searched for the production by considering both of
anomalous vertices $t-u-\gamma$ and $t-c-\gamma$ from the fact
that HERA has much higher sensitivity to $\kappa_{u\gamma }$ than
to $\kappa_{c\gamma }$, due to more favorable parton density
\cite{HC}.

In this paper we analyze anomalous single top production
considering both of the $\kappa_{q\gamma}$ and $\kappa_{qZ}$
couplings to see especially the contribution of the second
coupling. In order to make illustration clear we present the
numerical results for HERA and THERA \cite{THERA} colliders with
the CM energies $\sqrt{S}=320$ and  $1000$ $GeV$ respectively.
\section{Anomalous Top Decays}
We consider the most general effective Lagrangian proposed in
\cite{THJL} (generalized by the H1 Collaboration to include
$u\rightarrow t$ transition \footnote{This possibility is stated
also in ref. \cite{TC}}) to describe the anomalous top decays
$t\rightarrow q\gamma$ and $t\rightarrow qZ$ :
\begin{equation}
  \mathcal{L}_{eff}=\sum_{U=u,c}i\frac{ee_U}{\Lambda}\bar{t}\sigma_{\mu\nu}q^{\nu}
  \kappa_{\gamma,U}UA^{\mu}+\frac{g}{2\cos\theta_{W}}\bar{t}[\gamma_{\mu}(v_{Z,U}
  -a_{Z,U}\gamma^{5})+i\frac{1}{\Lambda}\sigma_{\mu,\nu}q^{\nu}\kappa_{Z,U}]UZ^{\mu}
  +h.c.,
\end{equation}

where $\sigma_{\mu,\nu}=(i/2)[\gamma^{\mu},\gamma^{\nu}]$,
$\theta_{W}$ is the Weinberg angle, $e_U$ is the electric charge
of up-type quarks, and $\Lambda$ denotes the scale up to which the
effective theory is assumed to hold. By convention we set
$\Lambda=m_t$ in the following.

We find:
\begin{equation}
 \Gamma(t\rightarrow
 q\gamma)=\alpha e_q^2\kappa_{q\gamma}^2m_{t}(1-\frac{m_q^2}{m_t^2})^3
\end{equation}

\begin{equation}
\Gamma(t\rightarrow qZ)=\frac{\alpha m_t}{4\sin^2\theta_{
W}\cos^2\theta_{
W}}(1-\frac{M_Z^2}{m_t^2})^2[\kappa_{qz}^2(1+\frac{M_Z^2}{2m_t^2})
-3v_q\kappa_{qz}+(a_q^2+v_q^2)(1+\frac{m_t^2}{M_Z^2})]
\end{equation}
where $\alpha$ is the fine structure constant, $M_Z$ is the mass
of $Z$ boson. Using the standard decay value $\Gamma(t\rightarrow
q W)=1.397$ $GeV$ and  the experimental constrains \cite{PDG}
\begin{center}
\begin{tabular}{c c c c c}
$BR(t\rightarrow q\gamma) < 3.2 \%$& & $CL$ &$95$& $\%$ \\
$BR(t\rightarrow q Z) < 33 \% $ & & $CL$ &$95$& $\%$
\end{tabular}
\end{center}
the upper limits on the couplings are easily obtained;
\begin{eqnarray}
\kappa_{q\gamma}< 0.24& &\kappa_{qZ}<0.27
\end{eqnarray}
In obtaining these values we assumed $Im$ $\kappa_{qZ}=0$ and
since $m_q\ll m_t$, we neglected the light quark masses.
\section{The Production of Single top quarks at HERA and THERA}
Differential production cross sections corresponding to the
diagrams mediated by $\gamma$, $Z$ and their interference term
respectively are;
\begin{eqnarray}\label{eq.4}
\frac{d\hat{\sigma}_{\gamma}}{d\hat{t}}&=&\frac{e_{0}^2e^4\kappa_{q\gamma}^2
  }{8\pi\hat{s}^2\hat{t}}[(2\hat{s}+\hat{t})-\frac{2\hat{s}(\hat{s}+\hat{t})}{m_t^2}-m_t^2]
 \\\nonumber\\
\frac{d\hat{\sigma}_Z}{d\hat{t}}&=&\frac{g_Z^{4}}{128\pi\hat{s}^2m_t^2
[(\hat{t}-M_Z^2)^2+\Gamma_Z^2M_Z^2]}\nonumber
\\
&&\{(a_e^2+v_e^2)\kappa_{qZ}^2[(2\hat{s}+
\hat{t})m_t^2-2\hat{s}(\hat{s}+\hat{t})-m_t^4]\hat{t}\nonumber
 \\
&&+2\kappa_{qZ}m_t^2[v_q(a_e^2+v_e^2)(\hat{t}-m_t^2)-2a_{e}v_{e}a_{q}(2\hat{s}
+\hat{t}-m_t^2)]\hat{t}
\\
&&+[(a_q^2+v_q^2)(a_e^2+v_e^2)[2\hat{s}^2
+2\hat{s}\hat{t}-(2\hat{s}+\hat{t})m_t^2+\hat{t}^2]
-4a_ev_ea_qv_q(2\hat{s}+\hat{t}-m_t^2)]m_t^2\}\nonumber
\\\nonumber\\
\frac{d\hat{\sigma}_{int}}{d\hat{t}}&=&\frac{g_Z^2e_0e^2\kappa_{q\gamma}(\hat{t}-M_Z^2)}
{16\pi\hat{s}^2m_t^2[(t-M_Z^2)^2+\Gamma_Z^2M_Z^2]}\nonumber
\\
&&\{a_ea_q(2\hat{s}+\hat{t}-m_t^2)m_t^2+v_e[2\hat{s}^2\kappa_{qZ}-(\hat{t}-m_t^2)(m_t^2(v_q+
\kappa_{qZ})-2\hat{s}\kappa_{qZ})]\}.
\end{eqnarray}
The total cross section is obtained by the
integral of the sum of these three :
\begin{equation}
  \sigma_{tot}=\int_{x_{min}}^{1}f_{q}(x)dx\int_{t_{-}}^{t_{+}}{\frac{d\hat{\sigma}}
  {d\hat{t}}}d\hat{t}
\end{equation}
where $t_{-}=-(\hat{s}-m_{t}^{2})$,$t_{+}=t_{cut}=-0.001 GeV$ and
$x_{min}=m_{t}^2/{S}$.

$f_q(x)$ is the quark distribution function inside the proton
\cite{MRST,GV}:
\begin{eqnarray*}
f_{u}(x)&=&A_ux^{\eta_1-1}(1-x)^{\eta_2}(1+\epsilon_u\sqrt{x}+\gamma_ux)
\\
f_{c}(x)&=&\frac{1}{2}N_5x^2[\frac{1}{3}(1-x)(1+10x+4x^2)+2x(1+x)\ln
x]  (n=2)
\end{eqnarray*}
where $A_u=0.8884$, $\eta_1=0.4710$, $\eta_2=3.404$,
$\epsilon_u=1.628$, $\gamma_u=9.628$ and $N_5=36$.

Figure ~\ref{fig.1}-~\ref{fig.3} display the photonic and total
cross-section due to up quark distribution as a function of the
center of mass energy (from HERA to THERA energy), taking,
$\kappa_{u\gamma}=0.3$, $0.2$, $0.1$ (solid lines) and
$\kappa_{uZ}=\kappa_{u\gamma}=0.3$, $0.2$, $0.1$ (dashed lines)
respectively. These values for the coupling constants were chosen
for purposes of demonstration only. Since the contribution of the
charm quark distribution to the production cross sections is much
less than that of the up distribution, instead of displaying the
corresponding figures, we tabulated the cross-section values in
Table ~\ref{tab.1}-~\ref{tab.3}. Table~\ref{tab.1} is formed by
considering only the photonic channel ($\kappa_{c\gamma}=0.1$,
$0.2$ and $0.3$).  In forming Table~\ref{tab.2} and~\ref{tab.3} we
considered both of the couplings $\kappa_{c\gamma}$ and
$\kappa_{cZ}$ with the values as indicated and therefore tabulated
the total cross section values at HERA and THERA colliders
respectively.
\section{Conclusions}
\begin{figure}
%\begin{minipage}[t]{0.5\linewidth}
  %\centering
  \includegraphics[width=13cm]{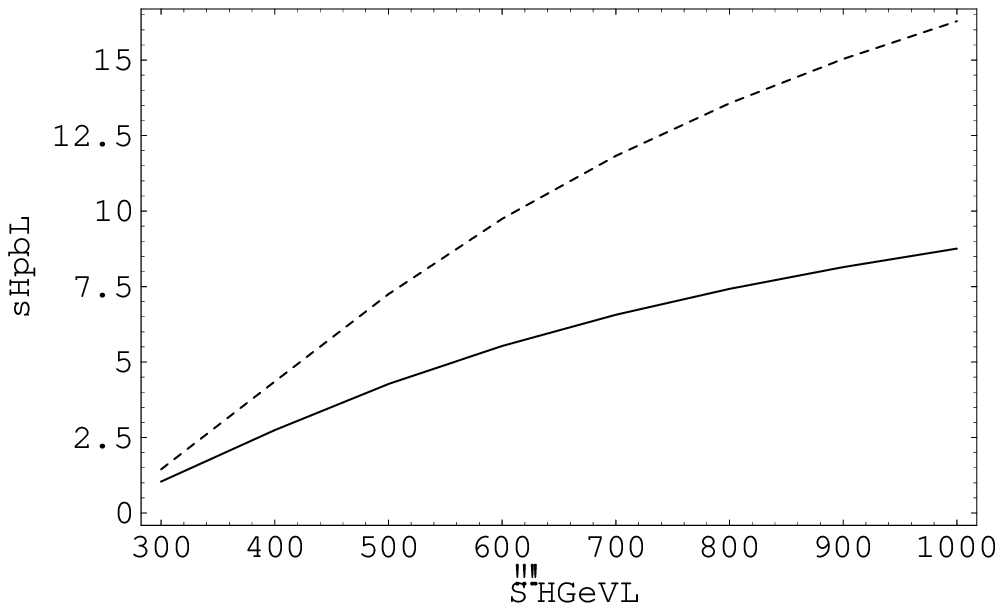}
  \caption{\label{fig.1}Photonic (solid line) and total (dashed line) production cross section for the FCNC single top
  quark as a function of  center of mass energy with $\kappa_{u\gamma}=0.3$ and $\kappa_{u\gamma}=\kappa_{uZ}=0.3$ respectively. }
 %\end{minipage}
\end{figure}

\begin{figure}
 %\begin{minipage}[t]{0.5\linewidth}
  %\centering
  \includegraphics[width=13cm]{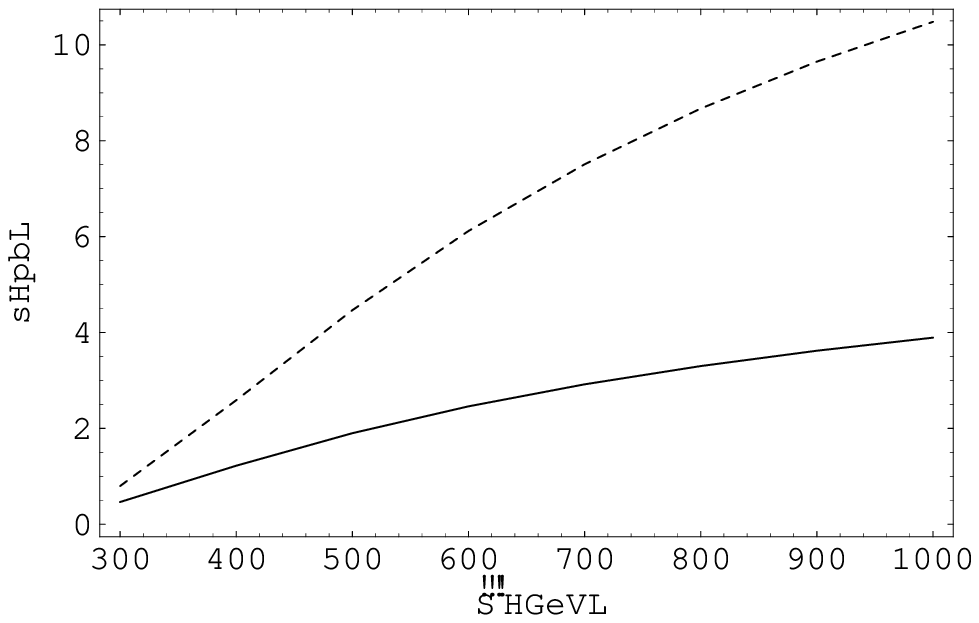}
  \caption{\label{fig.2}Photonic (solid line) and total (dashed line) production cross section for the FCNC single top
  quark as a function of center of mass energy with $\kappa_{u\gamma}=0.2$ and $\kappa_{u\gamma}=\kappa_{uZ}=0.2$ respectively. }
%\end{minipage}
\end{figure}

\begin{figure}
%\begin{minipage}[ b]{0.5\linewidth}
 %\centering
  \includegraphics[width=13cm]{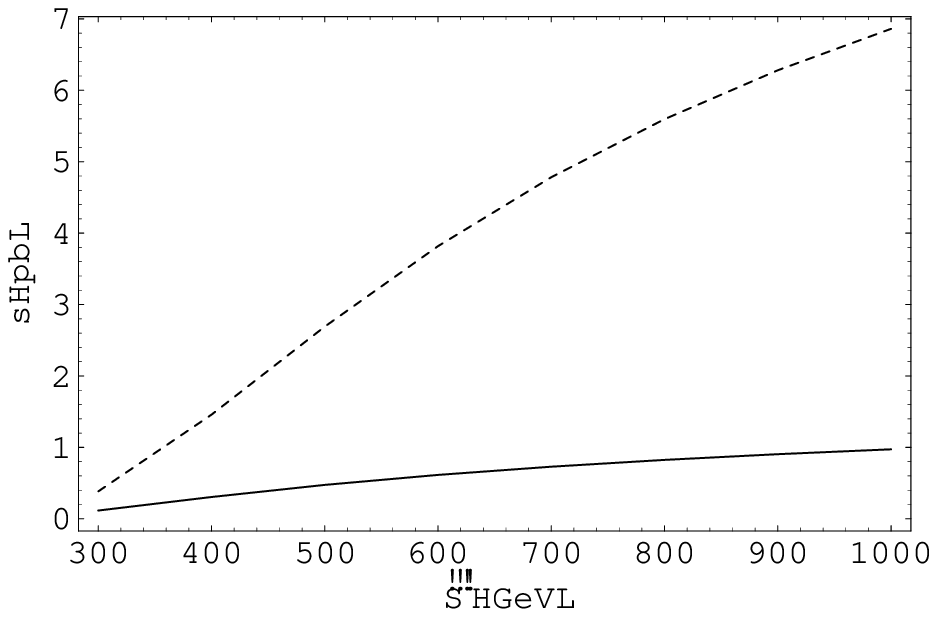}
  \caption{\label{fig.3}Photonic (solid line) and total (dashed line) production cross section for the FCNC single top
  quark as a function of center of mass energy with $\kappa_{u\gamma}=0.1$ and $\kappa_{u\gamma}=\kappa_{uZ}=0.1$ respectively.}
%\end{minipage}
\end{figure}

In most of the single top searches the dominant contribution to
the production is expected from the anomalous coupling
$\kappa_{q\gamma}$ only. However the contribution due to
$\kappa_{qZ}$ becomes important for the production with more
stringent bounds on these couplings. From Figure
~\ref{fig.1}-~\ref{fig.3} it is clear that contribution of the
anomalous production vertex $t-q-Z$ is negligible only at HERA
energies for $\kappa_{uV}=0.3$ (here $V$ refers to $\gamma$ and
$Z$ ). On the other hand the contribution of this vertex is
increased in two ways; one is by increasing the center of mass
energy and the other is as the couplings take more stringent
values. Figure ~\ref{fig.3} shows that if they take the same value
of $0.1$, the production is dominated by the anomalous vertex
$t-q-Z$ at THERA energies.
\begin{acknowledgements}
We would like to thank T.Tait for helpful discussion. This work
was supported by Abant Izzet Baysal University Research Fund.
\end{acknowledgements}

%\vspace{7cm}
\newpage
\begin{table}[h]
  \centering
\begin{tabular}{|c||c|c|c|}\hline
  $\kappa_{c\gamma }$ & 0.1 &0.2 & 0.3
  \\\hline\hline
  $\sqrt{S}=320$ & $1.6.10^{-3}$ & $6.2.10^{-3}$ & $14.0.10^{-3}$
  \\\hline
  $\sqrt{S}=1000$& $0.7.10^{-2}$ & $2.7.10^{-2}$ & $6.2.10^{-2}$  \\ \hline
\end{tabular}
  \caption{\label{tab.1}Single top production cross section in pb due to charm quark distribution mediated by photon }
\end{table}

\begin{table}[h]
  \centering
  \begin{tabular}{|c||c|c|c|c|}\hline
  &$\kappa_{qZ}=0$ &$\kappa_{cZ}=0.1$  &$\kappa_{cZ}=0.2$  &$\kappa_{cZ}=0.3$    \\\hline\hline
  $\kappa_{c\gamma}=0$  & $3.0.10^{-3}$ & $3.2.10^{-3}$ & $3.4.10^{-3}$& $3.7.10^{-3}$  \\\hline
  $\kappa_{c\gamma}=0.1$& $5.3.10^{-3}$ & $5.5.10^{-3}$ & $5.7.10^{-3}$ & $6.0.10^{-3}$  \\\hline
  $\kappa_{c\gamma}=0.2$& $10.6.10^{-3}$ & $10.8.10^{-3}$ & $11.0.10^{-3}$ & $11.3.10^{-3}$  \\\hline
  $\kappa_{c\gamma}=0.3$& $19.0.10^{-3}$ & $19.2.10^{-3}$ & $19.5.10^{-3}$& $19.8.10^{-3}$  \\\hline
  \end{tabular}
\caption{\label{tab.2}The total single top production cross
section in pb due to charm quark distribution at HERA }
\end{table}

\begin{table}[h]
  \centering
 \begin{tabular}{|c||c|c|c|c|}\hline
&$\kappa_{cZ}=0$ &$\kappa_{cZ}=0.1$  &$\kappa_{cZ}=0.2$
&$\kappa_{cZ}=0.3$    \\\hline\hline
  $\kappa_{c\gamma}=0$& $4.1.10^{-2}$ & $4.2.10^{-2}$  & $4.5.10^{-2}$  & $4.9.10^{-2}$  \\\hline
  $\kappa_{c\gamma}=0.1$& $5.0.10^{-2}$  & $5.1.10^{-2}$  & $5.4.10^{-2}$  &$5.8.10^{-2}$   \\\hline
  $\kappa_{c\gamma}=0.2$& $7.2.10^{-2}$ & $7.3.10^{-2}$  & $7.6.10^{-2}$  & $8.1.10^{-2}$  \\\hline
  $\kappa_{c\gamma}=0.3$& $10.7.10^{-2}$  & $10.9.10^{-2}$  & $11.2.10^{-2}$  & $11.7.10^{-2}$  \\\hline
  \end{tabular}
\caption{\label{tab.3}The total single top production cross
section in pb due to charm quark distribution at THERA}
\end{table}
\vspace{7cm}
\vspace{15cm}
\end{document}